\documentclass[preprint,amsmath,amssymb]{revtex4}
\usepackage{graphicx}
\usepackage{tabularx}
\usepackage{makecell}
\usepackage{dcolumn}
\usepackage{bm}

\begin{document}
\title{Static and stationary dark fluid universes: A gravitoelectromagnetic perspective}

\author{${\rm M.\;Nouri}$-${\rm Zonoz}^{\;(a)}$\footnote{Electronic
address:~nouri@ut.ac.ir \; (Corresponding author)} and ${\rm A.\;Nouri}$-${\rm Zonoz}^{\;(b)}$\footnote {Electronic
address:~ahmad.nouri.zonn@gmail.com } }
\affiliation{(a): Department of Physics, University of Tehran, North Karegar Ave., Tehran 14395-547, Iran.\\ 
(b): Department of Physics, Shahid Beheshti University, G.C., Evin, Tehran, 19839, Iran.}

\begin{abstract}
We introduce a physical characterization of the static and stationary perfect fluid solutions of the Einstein field equations 
with a single or 2-component perfect fluid sources, according to their gravitoelectric and gravitomagnetic fields.
The absence or presence of either or both of these fields could restrict the equations of state 
of the underlying perfect fluid sources. 
As the representative of each family of solutions, we consider those spaces that include the cosmological term as a dark fluid source with 
the equation of state $p=-\rho = constant$. 
\end{abstract}
\maketitle

\section{Introduction and motivation} 
There are detailed discussions of exact solutions of Einstein field equations (EFEs), and their characterization based on different symmetry groups of 
either geometric objects, such as Weyl and Ricci tensors, or the energy-momentum tensor of the source \cite{Exact1}. Static and stationary perfect fluid solutions, 
on the other hand have played  a pivotal role in the evolution of the cosmological models, and have been 
discussed extensively in the exact solution literature \cite{Exact1, Exact2}.
In these solutions there could be more than one perfect fluid 
source, each with a different barotropic equation of state (EOS).
Employing the quasi-Maxwell form of the Einstein field equations for multi-component perfect fluid sources, 
here we show how a combination of different choices for the gravitoelectromagnetic (GEM) fields, along with different EOS for different perfect fluid sources, 
could naturally lead to well-known static and stationary perfect fluid spacetimes as the representative of each class, hence furnishing a physical characterization 
of these spacetimes. 
The presence or absence of either or both of the gravitoelectric (GE) and gravitomagnetic (GM)
fields could in some cases, not only restrict the minimum number of the perfect fluid sources, but also fix their EOS. 
We will treat the cosmological term, $\Lambda g_{ab}$, as a perfect (dark) 
fluid source with EOS $p_\Lambda = -\rho_\Lambda$, in which $\rho_{\Lambda} = \frac{\Lambda}{8\pi}$.
Interestingly enough we will find out that in some cases the sign of the cosmological constant, or equivalently $\rho_\Lambda$, is fixed by our choice of the GEM fields.  
Indeed, as an interesting example of the above characterization, it has already been shown that the de Sitter space, and the so called de Sitter-type 
spacetimes are the only  {\it static single-component} perfect fluid solutions of EFE in the {\it non-comoving} frames \cite{NKR}. Characterizing them in this way, 
the apparent paradox raised by some authors \cite{Rindler, GP} on why there are different static spacetimes with $\Lambda$ as their only parameter was resolved.
De Sitter-type spacetimes are  axially  and cylindrically symmetric static {\it Einstein spaces} (solutions of $R_{ab} = \Lambda g_{ab}$) with $\Lambda$ as their only parameter, 
so that they were first expected to be the good old de Sitter spacetime just in different coordinate systems.
But they were found to be genuinely different from de Sitter space,  when their curvature invariants as well as their dynamical forms in the 
{\it comoving synchronous} coordinate systems were calculated. 
These findings motivated the idea that one should consider a {\it perfect fluid nature} for the cosmological term and assign a 4-velocity to this dark fluid, in order 
to be able to interpret the {\it directional expansion} of the de Sitter-type spacetimes in their dynamical forms \cite{NKR}.\\
Here we will show how the static and stationary dark fluid universes could be characterized in terms of their gravitoelectromagnetic fields in a fundamental observer's 
frame adapted to the time-like Killing vector field of the corresponding spacetimes.\\
The outline of the paper is as follows. In the next section we introduce the threading formulation of spacetime decomposition, 
and the quasi-Maxwell form of the Einstein field equations. In four subsections of section III, using the characterization based on the quasi-Maxwell form of EFE, and the
gravitoelectromagnetic fields, 
we show how the  homogeneous static and stationary perfect fluid solutions could be categorized.\\
Throughout, the Latin indices run from 0 to 3 while the Greek ones run from 1 to 3, and we will use the units in which $c= G=1$.
%%%%%%%%%%%%%%%%%%%%%%%%%%%%%%%%%%%%%%%%%%%%%%%%%%
\section{Gravitoelectromagnetism and the quasi-Maxwell form of the Einstein field equations}
The $1+3$ or threading formulation of spacetime decomposition is the decomposition of spacetime by the worldlines of {\it fundamental observers} who are at fixed spatial 
points in a gravitational field. In other words, these worldlines, sweeping the history of the spatial positions of the fundamental observers, 
decompose the underlying spacetime into timelike threads  \cite{Lan}. In stationary asymptotically flat spacetimes, these observers 
are at rest with respect to the distant observers
in the asymptotically flat region. Employing propagation  of radar signals between two nearby fundamental observers  
the spacetime metric could be expressed in the following general form,
\begin{equation}\label{ds0}
d{s^2} = d\tau_{sy}^2 - d{{l}^2} = {g_{00}}{(d{x^0} - {g_\alpha }d{x^\alpha })^2} - {{\gamma}_{\alpha \beta }}d{x^\alpha }d{x^\beta },
\end{equation}
where ${g_\alpha } =  - \frac{{{g_{0\alpha }}}}{{{g_{00}}}}$ and
\begin{equation}\label{gamma0}
{{\gamma} _{\alpha\beta}} =  - {g_{\alpha \beta }} + \frac{{{g_{0\alpha }}{g_{0\beta }}}}{{{g_{00}}}} \;\; ; \;\;  {{\gamma} ^{\alpha \beta }} =  - {g^{\alpha \beta }},
\end{equation}
is the spatial metric of a 3-space $\Sigma_3$, on which $d{{l}}$ gives the element of spatial distance between any two nearby events.
Also, $d{\tau _{sy}} = \sqrt {{g_{00}}} (d{x^0} - {g_\alpha }d{x^\alpha })$ gives the infinitesimal interval of the so-called {\it synchronized proper time} 
between any two events. In other words any two simultaneous events have a world-time difference of $d x^0 = {g}_\alpha dx^\alpha$. 
The origin of this definition of a time interval could be explained through the following procedure for definition of a prticle's 3-velocity. 
If the particle passes point B (with spatial coordinates $x^\alpha$) at the 
moment of world time ${x^0}$ and arrives at the infinitesimally distant point A (with spatial coordinates $x^\alpha + d x^\alpha$) at the moment ${x^0} + d{x^0}$, 
then to determine its velocity we 
must now take, difference between ${x^0} + d{x^0}$ and the moment ${x^0} - \frac{{{g_{0\alpha }}}}{{{g_{00}}}}d{x^\alpha }$ which is {\it simultaneous} at the point 
B with the moment ${x^0}$ at the 
point A (Fig. 1). Now  upon dividing the infinitesimal spatial coordinate interval $dx^\alpha$ by this time difference the 3-velocity of a particle in the underlying 
spacetime is given by \cite{Lan, MN}
\begin{equation}\label{velo}
{{v}^\alpha } = \frac{{d{x^\alpha }}}{{d{\tau _{sy}}}} = \frac{{d{x^\alpha }}}{{\sqrt {{g_{00}}} (d{x^0} - {g_\alpha }d{x^\alpha })}}.
\end{equation}
%%%%%%%%%%%%%%%%%%%%%%%%%%%%%%%%%%%%%%%%%%%%%%%%%%%%%%%%%%%%%%%%%%%%%
\begin{figure}\label{1}
\includegraphics[scale=0.60]{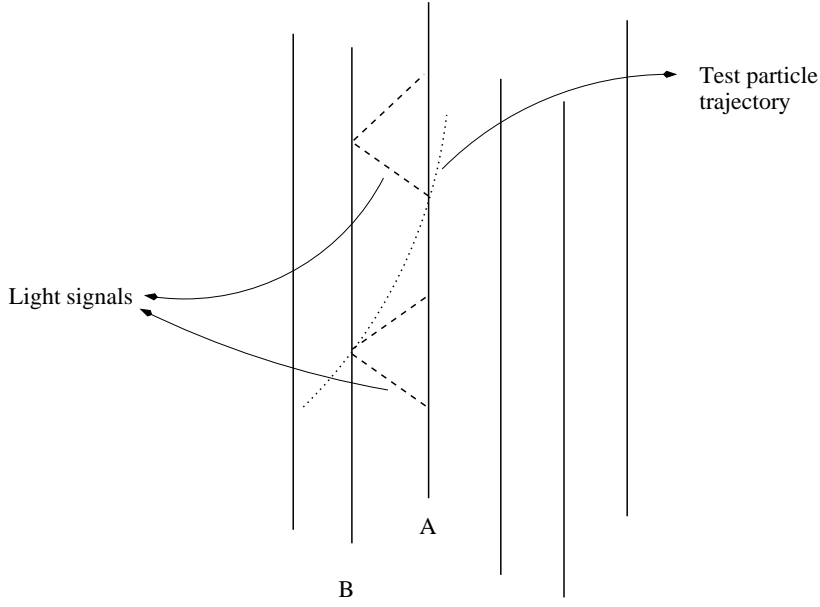}
\caption{A congruence of nearby worldlines of fundamental observers and a test particle crossing them. The observers {\bf A} and {\bf B} exchange radar signals
to define spatial distances and the 3-velocity of a test particle in terms of the synchronized proper time.}
\end{figure}
%%%%%%%%%%%%%%%%%%%%%%%%%%%%%%%%%%%%%%%%%%%%%%%%%%%%%%%%%%%%%%%%%%%%%%555
Obviously, in the case of static spacetimes (i.e., $ g_{0\alpha} = 0$) the above definition 
reduces to the proper velocity defined by $v^\alpha = \frac{1}{\sqrt{g_{00}}}\frac{dx^\alpha} {dx^0}$ \footnote{For a detailed discussion on the definition of 3-velocity 
refer to \cite{Ghare}.}. \\
Substituting the above definition of 3-velocity in Eq. \eqref{ds0}, one can show the following relation between the proper and synchronized proper times 
\begin{equation}\label{ds00}
d{\tau^2} = {g_{00}}{(dx^0 - g_{\alpha }d{x^\alpha })^2}[1 - {{v}^2}] = d\tau_{sy.}^2 (1-{v}^2).
\end{equation}
Also the components of the 4-velocity $u^i =dx^{i}/d\tau$ of a test particle, in terms of the components of its 3-velocity, are given by
\begin{eqnarray}\label{4u}
u^\alpha = \frac{v^\alpha}{\sqrt{1-v^2}}, \;\; u^0 = \frac{1}{\sqrt{1-v^2}}\left( \frac{1}{\sqrt{g_{00}}}+g_{\alpha}v^{\alpha}\right). 
\end{eqnarray} 
Obviously the comoving frame is defined by $v^{\alpha}=0$ leading to $u^i = (\frac{1}{\sqrt{g_{00}}}, 0, 0, 0)$ as expected. \\
Applying the above formalism we define the 3-force acting on a test particle in a stationary gravitational field as the 3-dimensional  covariant derivative of 
the particle's 3-momentum with respect to the synchronized proper time \cite{Lan,MN}, i.e,
\begin{equation}\label{force01}
{f}^\mu \equiv  \frac{D p^\mu}{d\tau_{sy}} = \sqrt{1-v^2} \frac{D p^\mu}{d\tau},
\end{equation}
in which we used equation \eqref{ds00} to write it in terms of the proper time. Since by definition $p^\mu=mu^\mu$, we use the spatial components of the geodesic 
equation for a test particle, namely 
\begin{eqnarray}\label{geodesic}
\frac{du^\mu}{d\tau} = -\Gamma^\mu_{ab} u^a u^b = - \Gamma^\mu_{00} (u^0)^2  - 2\Gamma^\mu_{0\beta} u^0 u^{\beta} -\Gamma^\mu_{\alpha\beta} u^\alpha u^\beta .
\end{eqnarray}
and substitute expressions for the connection coefficients in terms of the 3-dimensional objects and the 4-velocity components from \eqref{4u}, to arrive at
the following expression for the  gravitational Lorentz-type 3-force,
\begin{equation}\label{force1}
{f}^\mu = {\sqrt{1-v^2}} \frac{d}{d\tau}\frac{mv^\mu}{\sqrt{1-v^2}} + \lambda^{\mu}_{\alpha\beta} \frac{mv^\alpha v^\beta}{\sqrt{1-v^2}},
\end{equation}
in which $\lambda^{\mu}_{\alpha\beta}$ is the 3-dimensional Christoffel symbol constructed from ${{\gamma} _{\alpha\beta}}$. Intuitively, this shows that test particles
moving on the geodesics of a stationary spacetime depart from the geodesics of the 3-space $\Sigma_3$ as if acted on by the above-defined gravitational 3-force.
Lowering the index, in its vectorial form the above expression could be written in the following form,
\begin{equation}\label{force}
{\bf f}_g =  \frac{m}{\sqrt{1-{v^2}}}\left( {\bf E}_g + {\rm {\bf v}}\times \sqrt{g_{00}}{\bf B}_g\right),
\end{equation}
in which the gravitoelectric (GE) and gravitomagnetic (GM) 3-fields (with lower and upper indices respectively), are defined as follows 
\footnote{We note that the differential
operations in these relations are defined in the 3-space $\Sigma_3$ with metric $\gamma_{\mu\nu}$. Specifically, divergence and curl of a vector are defined as 
${\rm div} \;\textbf{V}=\frac{1}{\sqrt{\gamma}}~\frac{\partial}{\partial{x^i}}(\sqrt{\gamma}~V^i)~~~{\rm and} ~~~({\rm curl}\; \textbf{V})^i=\frac{1}{2\sqrt{\gamma}}~\epsilon^{ijk}
(\frac{\partial{V_k}}{\partial{x^j}}-\frac{\partial{V_j}}{\partial{x^k}}),$ respectively
with $\gamma=det~\gamma_{ij}$.}
\begin{gather}
\textbf{E}_g = -{\bf {\nabla}} \ln \sqrt{h} \;\;  ; \;\; (h \equiv  g_{00})\label{pot}\\
\textbf{B}_g = curl~({\bf A}_g)\;\;  ; \;\; ({A_g}_\alpha \equiv g_{\alpha})\label{bg},
\end{gather}
in which $\ln \sqrt{h}$ and ${\bf A}_g$  are the so-called GE and GM potentials respectively \cite{NP}. We notice that GE part of the 
gravitoelectromagnetic (GEM) Lorentz force \eqref{force} is the general relativistic version of the gravitational force in Newtonian gravity \cite{Nouri18}, while its GM 
part has no counterpart in Newtonian gravity. Obviously by their definition, they satisfy the following constraints
\begin{equation}
\nabla \times~\textbf{E}_g=0, ~~~\nabla \cdot  \textbf{B}_g=0. 
\end{equation}
Now in terms of the GEM fields measured by the fundamental observers, the
Einstein field equations for a multi-component fluid sources, each having an energy-momentum tensor $T_{ab}=(p + \rho)u_a u_b - p g_{ab}$ with $u^a u_a = 1$,
could be written in the following quasi-Maxwell form \cite{MN},
\begin{gather} 
\nabla \cdot \textbf{E}_g= \frac{1}{2} h B^2_g+E^2_g - {8\pi}\Sigma_i \left(\dfrac{p_i+\rho_i}{1-{v_i}^2}-\dfrac{\rho_i-p_i}{2}\right) \label{r00} \\  
\nabla \times  (\sqrt{h}\textbf{B}_g)=2 \textbf{E}_g \times (\sqrt{h}\textbf{B}_g)-{16\pi}\Sigma_i\left(\dfrac{p_i+\rho_i}{1-{v_i}^2}\right) {\textbf{v}_i} \label{r01} \\
{^{(3)}}P^{\mu\nu}=-{E}_g^{\mu;\nu}+\frac{1}{2}h(B_g^\mu B_g^\nu - B_g^2 \gamma^{\mu\nu})+ {E}_g^\mu E_g^\nu+ 
{8\pi}\Sigma_i \left(\dfrac{p_i+\rho_i}{1 - {v_i}^2}{v_i}^\mu {v_i}^\nu+\dfrac{\rho_i-p_i}{2}\gamma^{\mu\nu}\right) \label{3ricci},
\end{gather}
in which ${\bf v}_i$ is the 3-velocity of the $i$-th component of the source fluid as defined in (\ref{velo}). Also ${^{(3)}}P^{\mu\nu} $ is the three-dimensional Ricci 
tensor made out of the 3-d metric $\gamma^{\mu\nu}$. Here we focus on  2-component fluid sources so that  ${i=1,2}$.\\
The above formalism has been employed to derive gravitational analogs of some  well known electromagnetic effects \cite{MN, Nouriz, Fil}. It has also been used to 
discover and interpret exact solutions of the EFEs and study gravitational lensing \cite{Gemex}.
%%%%%%%%%%%%%%%%%%%%%%%%%%% %%%%%%%%%%%%%%%%%%%%%%%%%%%%%%%%%%%%%%%%%%%%%%%%%%%%%%%%
\section{Static and stationary perfect fluid solutions}
Using the quasi-Maxwell form of the Einstein field equations \eqref{r00}-\eqref{3ricci}, in what follows we will employ the following three criteria to characterize 
well-known static and stationary perfect fluid solutions:\\
I-Vanishing of either or both of the gravitoelectric ($E_g$) and gravitomagnetic ($B_g$) fields. \\
II-Number of perfect fluid components and their corresponding EOS.\\
III-Fluid components and their frames: either a comoving frame or a non-comoving one.\\
Indeed in what follows we will find out that applying the first criterion to Eqs. \eqref{r00} and \eqref{r01}, will automatically restrict 
both the minimum number of the fluid components as well as their EOS in a given frame. \\
As the repesentative solution in each family with the lowest number of parameters, in the case of static spacetimes we consider spherically symmetric
solutions whereas in the case of stationary spacetimes we restrict our attention to axially and cylindrically symmetric cases.
%%%%%%%%%%%%%%%%%%%%%%%%%%%%%%%%%%%%%%%%%%%%%%%%%%%%%%%%%%%%%%%%%%%%%5
\subsection{Spacetimes without gravitoelectromagnetic fields $E_g$ and $B_g$: Einstein Static Universe}
Substituting $E_g=0$ and $B_g=0$ in Eqs. \eqref{r00}-\eqref{r01} we end up with the following equations, 
\begin{gather} 
\Sigma_i \left(\dfrac{p_i+\rho_i}{1-{v_i}^2}-\dfrac{\rho_i-p_i}{2}\right) = 0 \label{r001} \\  
\Sigma_i\left(\dfrac{p_i+\rho_i}{1-{v_i}^2}\right) {\textbf{v}_i} = 0 \label{r011} \\
{^{(3)}}P^{\mu\nu}= {8\pi}\Sigma_i \left(\dfrac{p_i+\rho_i}{1 - {v_i}^2}{v_i}^\mu {v_i}^\nu+
\dfrac{\rho_i-p_i}{2}\gamma^{\mu\nu}\right) \label{3ricci1}.
\end{gather}
We notice that the first two equations only include the source specifications and any solution has the following characteristics:\\ 
1-It is a static spacetime.\\
2-With $E_g=B_g=0$ in the GEM Lorentz force \eqref{force}, there will be no gravitational force acting on test particles in this spacetime, i.e. 
particles stay where they are.\\
Now Eq. \eqref{r011}  seems to be  satisfied for a {\it single} component perfect fluid, either with \\ 
A-any EOS in a {\it comoving frame} ($v=0$) or B- a dark fluid with EOS  $p= - \rho$.\\ 
If we take the first case and substitute $v=0$ in Eq. \eqref{r001}, that will fix the fluid EOS to $p=\rho/3$ which is that of {\it incoherent 
radiation}. Of course photons as particles of radiation are not timelike and do not satisfy $u^a u_a = 1$. 
Now if we choose the second single component fluid with EOS $p= - \rho$, that will not satisfy Eq. \eqref{r001}. 
Also it is noticed that we have found these results without recourse to the last equation and in fact none of these choices satisfy Eq. \eqref{3ricci1} which 
takes the forms ${^{(3)}}P^{\mu\nu}= \pm 8\pi p \gamma^{\mu\nu}$ (with the minus sign for the dark fluid) for a constant pressure. \\
From a physical point of view, that a single-component fluid does not lead to a solution is expected, 
since any kind of normal matter will produce attractive gravity, and hence leads to a collapsing system with $F_g \neq 0$, 
hence contradicting the second point above. Indeed this was the problem Einstein faced in his 1917 effort to find an static Universe. \\
Therefore to have a solution  we need at least a 2-component fluid which, when plugged into Eqs. \eqref{r001}-\eqref{3ricci1}, leads to the following equations;
\begin{gather}
\left(\dfrac{p_1+\rho_1}{1-{v_1}^2}-\dfrac{\rho_1-p_1}{2}\right) + \left(\dfrac{p_2+\rho_2}{1-{v_2}^2}-\dfrac{\rho_2-p_2}{2}\right) = 0 \label{r002} \\  
\left(\dfrac{p_1+\rho_1}{1-{v_1}^2}\right) {\textbf{v}_1} + \left(\dfrac{p_2+\rho_2}{1-{v_2}^2}\right) {\textbf{v}_2}  = 0 \label{r012} \\
{^{(3)}}P^{\mu\nu}= {8\pi} \left(\dfrac{p_1+\rho_1}{1 - {v_1}^2}{v_1}^\mu {v_1}^\nu +
\dfrac{\rho_1-p_1}{2}\gamma^{\mu\nu}\right) + {8\pi} \left(\dfrac{p_2+\rho_2}{1 - {v_2}^2}{v_2}^\mu {v_2}^\nu +
\dfrac{\rho_2-p_2}{2}\gamma^{\mu\nu}\right) \label{3ricci2}.
\end{gather}
Looking at Eq. \eqref{r012}, we notice that one can always satisfy it by choosing one of the fluid components (with any well-known EOS) to be 
in the {\it comoving frame} (say $v_1=0$), and the second component to have an EOS $p_2 = -\rho_2$, that of a dark fluid. Obviously the next step is to 
put these values in Eq. \eqref{r002} to find the
relation between the two component densities (or pressures). The last equation, Eq. \eqref{3ricci2} serves for the application of 
the required symmetry. Now we could have for the fluid
in the comoving frame either  1- dust ($p=0$), 2-radiation ($p=\rho /3$) or 3- stiff matter ($p=\rho$) leading respectively to:\\
1-Einstein static universe in which the relation between the two fluid densities is given by $\rho_{\Lambda} =\frac{ \rho_{dust}}{2}$ or 
equivalently $\Lambda = 4\pi \rho_{dust}$.\\
2-Static universe filled with incoherent radiation in which the relation between the two fluid densities is given by $\rho_{\Lambda} = \rho_{radiation}$ or equivalently 
$\Lambda = 8\pi \rho_{radiation}$ \footnote{We notice that this case could be treated in the present formalism, 
if we consider massive relativistic particles as incoherent radiation.}.\\
3-Static universe filled with stiff matter (SM) in which the relation between the two fluid densities is given by $\rho_{\Lambda} = 2 \rho_{SM}$ or equivalently 
$\Lambda = 16\pi \rho_{SM}$.\\
In terms of the cosmological constant, the metric of the above three static spherically symmetric spacetimes are given by,
\begin{equation}\label{ds03300}
d{s^2} = {dt}^2 -\frac{dr^2}{1 -  \frac{\Lambda}{\beta}r^2} - r^2  (d\theta^2 + \sin^2 \theta d\phi^2),
\end{equation}
in which $\beta = 1, 3/2, 2$ for dust, radiation and stiff matter sources respectively. The above form of the metric shows clearly the flat space limit $\Lambda \rightarrow 0$, 
and the obvious fact that $\frac{\Lambda}{\beta}$ gives the spacetime curvature for different values of ${\beta}$.
In summary, vanishing of both gravitoelectric and gravitomagnetic fields ($E_g=B_g=0$)  is consistent with the static nature of 
this solution where the repulsion of the dark fluid counterbalances the attraction of the non-dark element which could be dust, incoherent radiation or stiff matter. 
%%%%%%%%%%%%%%%%%%%%%%%%%%%%%%%%%%%%%%%%%%%%%%%%%%%%%%%%%%%%%%%%%%%%%%%%%
\subsection{Spacetimes without a gravitomagnetic field $B_g$: de Sitter spacetime}
Starting from Eqs. \eqref{r00}-\eqref{3ricci} and setting $B_g=0$, we end up with the following equations;
\begin{gather} 
\nabla \cdot \textbf{E}_g= E^2_g - {8\pi}\Sigma_i \left(\dfrac{p_i+\rho_i}{1-{v_i}^2}-\dfrac{\rho_i-p_i}{2}\right) \label{r003} \\  
\Sigma_i\left(\dfrac{p_i+\rho_i}{1-{v_i}^2}\right) {\textbf{v}_i} = 0 \label{r013} \\
{^{(3)}}P^{\mu\nu}=-{E}_g^{\mu;\nu} + {E}_g^\mu E_g^\nu+ 
{8\pi}\Sigma_i \left(\dfrac{p_i+\rho_i}{1 - {v_i}^2}{v_i}^\mu {v_i}^\nu+\dfrac{\rho_i-p_i}{2}\gamma^{\mu\nu}\right) \label{3ricci3}.
\end{gather}
Again looking at Eq. \eqref{r013}, it seems that we could have a one-component fluid solution either with any EOS in a comoving frame, or if we are looking for a solution
in a non-comoving frame, then the only choice would be a dark fluid, namely $p = -\rho$, but now, unlike the previous case in the last section, such a choice is not 
forbidden by the other two equations. Indeed this case has been thoroughly discussed in \cite{NKR}, where it is shown that it leads to a unique characterization of de Sitter 
and  {\it de Sitter-type} spacetimes as the only one-component {\it static} perfect fluid solutions of Einstein field equations in a {\it non-comoving frame}. 
The well known de sitter spacetime
\begin{equation}\label{desit1}
ds^2 = (1-\frac{\Lambda r^2}{3}) c^2 dt^2 - (1-\frac{\Lambda r^2}{3})^{-1} dr^2 - r^2 (d\theta^2 + \sin^2 \theta d\phi^2),
\end{equation}
is the spherically symmetric member of this family, and indeed their representative, which could be easily shown to satisfy Eqs. \eqref{r003} and \eqref{3ricci3}. The axially and 
cylindrically symmetric members of the same family are given by \cite{Nariai,bonn,Rindler}
\begin{equation}\label{desit3-2}
ds^2 = (1-{\Lambda z^2}) c^2 dt^2 - (1-{\Lambda z^2})^{-1} dz^2 - \frac{1}{(1+\frac{\Lambda}{4} {{\rho}}^2)^2} ({d{{\rho}}^2} +{{\rho}}^2 d\phi^2),
\end{equation}
and
\begin{equation}\label{cyldes}
 ds^2={\cos^{4/3}\bigg(\frac{\sqrt{3\Lambda}}{2}\rho\bigg) (d t^2
 - d z^2) - d\rho^2 - \frac {4}{3\Lambda}
 \sin^2\bigg(\frac{\sqrt{3\Lambda}}{2}\rho\bigg)
 \cos^{-2/3}\bigg(\frac{\sqrt{3\Lambda}}{2}\rho\bigg) d\phi^2},
 \end{equation}
respectively.  It should be noted that the same approach could also be applied to
dark fluids with $\rho_\Lambda < 0$, leading to the anti-de Sitter spacetime and its axially and cylindrically symmetric counterparts \cite{Linet,Bonnor}.
Obviously apart from these 1-parameter solutions there are other solutions of \eqref{r003}-\eqref{3ricci3} with two or more parameters. 
The simplest 2-parameter solution is the well-known Schwarzschild-de Sitter space which includes the mass parameter.
%%%%%%%%%%%%%%%%%%%%%%%%%%%%%%%%%%%%%%%%%%%%%%%%%%%%%%%%%%%%%%%%%%%%%%%%%%%%%%%
\subsection{Spacetimes without a gravitoelectric field $E_g$: The G\"{o}del Universe}
Spacetimes with a gravitomagnetic field are stationary spacetimes and the {\it absence} of the gravitoelectric field requires a constant time-time component of the metric,
i.e $h \equiv a^2 = constant$. 
Looking for cylindrically symmetric solutions \footnote{For a recent review on cylindrical gravitational fields refer to \cite{Bron}.}, these observations reduce the 
general form of the metric (in a cylindrically symmetric coordinate system) into \cite{Exact1},
\begin{equation}\label{ds02}
d{s^2} = a^2 [{dt} + A(r)d\phi]^2 - d\rho^2 - e^{2K(r)} dz^2 - G(r) d\phi^2,,
\end{equation}
which has a gravitomagnetic field along {\it the z-axis}. Starting from Eqs. \eqref{r00}-\eqref{3ricci} and setting $E_g=0$, 
we end up with the following equations;
\begin{gather} 
\frac{1}{2} a^2 B^2_g = {8\pi}\Sigma_i \left(\dfrac{p_i+\rho_i}{1-{v_i}^2}-\dfrac{\rho_i-p_i}{2}\right) \label{r004} \\  
a \nabla \times  (\textbf{B}_g) = -{16\pi}\Sigma_i\left(\dfrac{p_i+\rho_i}{1-{v_i}^2}\right) {\textbf{v}_i} \label{r014} \\
{^{(3)}}P^{\mu\nu}=\frac{1}{2}a^2(B_g^\mu B_g^\nu - B_g^2 \gamma^{\mu\nu})+  
{8\pi}\Sigma_i \left(\dfrac{p_i+\rho_i}{1 - {v_i}^2}{v_i}^\mu {v_i}^\nu+\dfrac{\rho_i-p_i}{2}\gamma^{\mu\nu}\right) \label{3ricci4}.
\end{gather}
Lets try a single perfect fluid source with any linear barotropic EOS with constant pressure (density), excluding that of a dark-type ($p = - \rho = constant$), then  
Eqs. \eqref{r004} and \eqref{r014} are simultaneously satisfied, {\it only} in a {\it comoving} frame ($v=0$), leading to a {\it uniform} and curl-free gravitomagnetic field. 
This includes for example stiff matter ($p = \rho = constant $), which 
when plugged into \eqref{r014} leads to the following equations
\begin{gather} 
{B_g}^2 = 32 \frac{\pi}{a^2}\rho_{SM} \label{r0041} \\  
\nabla \times  (\textbf{B}_g) = 0 \label{r0141} \\
{^{(3)}}P^{\mu\nu}=\frac{1}{2}a^2(B_g^\mu B_g^\nu - B_g^2 \gamma^{\mu\nu}) \label{3ricci41},
\end{gather}
in which, as mentioned, the first two equations refer to a {\it uniform} gravitomagnetic field.
Indeed this form of a source matter, satisfying the last 
equation \eqref{3ricci41}, will result in a solution which 
is the famous G\"{o}del universe \cite{godel} in which the source of the spacetime is stiff matter in a comoving frame.\\
The one-component perfect fluid of the dark-type with EOS $p = - \rho = constant$, although satisfying Eqs. \eqref{r004} and \eqref{r014} for $\rho_\Lambda < 0$ ($\Lambda < 0$), 
is excluded as it will not lead to a solution of \eqref{3ricci4} which will take the 
form ${^{(3)}}P^{\mu\nu}=\frac{1}{2}a^2 B_g^\mu B_g^\nu + 16 \pi \rho_\Lambda \gamma^{\mu\nu}$ \footnote{Indeed one could show that equations for ${^{(3)}}P^{\rho\rho}$  
and ${^{(3)}}P^{zz}$ lead to $\rho_\Lambda = 0$.}.\\
If on the other hand we insist on having a dark fluid component, as we have done so far, then we should look for a solution of the above equations 
with  two perfect fluid sources namely, 
\begin{gather} 
\frac{1}{2} a^2 B^2_g = {8\pi}\left( (\dfrac{p_1+\rho_1}{1-{v_1}^2}-\dfrac{\rho_1-p_1}{2}) + (\dfrac{p_2+\rho_2}{1-{v_2}^2}-\dfrac{\rho_2-p_2}{2})  \right) \label{r005} \\  
a \nabla \times  (\textbf{B}_g) = -{16\pi} \left( \dfrac{p_1+\rho_1}{1-{v_1}^2}  {\textbf{v}_1} + \dfrac{p_2+\rho_2}{1-{v_2}^2}  {\textbf{v}_2} \right)\label{r015} \\
{^{(3)}}P^{\mu\nu}=\frac{1}{2}a^2(B_g^\mu B_g^\nu - B_g^2 \gamma^{\mu\nu}) + \nonumber \\ 
{8\pi} \left((\dfrac{p_1+\rho_1}{1 - {v_1}^2}{v_1}^\mu {v_1}^\nu+\dfrac{\rho_1-p_1}{2}\gamma^{\mu\nu})
+ (\dfrac{p_2+\rho_2}{1 - {v_2}^2}{v_2}^\mu {v_2}^\nu + \dfrac{\rho_2-p_2}{2}\gamma^{\mu\nu}) \right) \label{3ricci5}.
\end{gather}
To have a curl-free gravitomagnetic field, Eq. \eqref{r015} invite us to choose, as in the case of the 
static universes discussed in section {\bf III-A}, a dust component \footnote{we note that unlike the case of static universes, here we are not allowed to choose incoherent radiation, 
as it will not be consistent with the  cylindrical symmetry.} in the comoving frame, plus a  dark component ($p = -\rho$). These two sources substituted in the above equations
lead to,
\begin{gather} 
{B_g}^2 = 16 \frac{\pi}{a^2}(\frac{\rho_{dust}}{2} - \rho_{\Lambda}) \label{r0051} \\  
\nabla \times  (\textbf{B}_g) = 0 \label{r0151} \\
{^{(3)}}P^{\mu\nu}=\frac{1}{2}a^2(B_g^\mu B_g^\nu - B_g^2 \gamma^{\mu\nu}) + {8\pi} \gamma^{\mu\nu} (\frac{\rho_{dust}}{2} + \rho_{\Lambda}) \label{3ricci51}.
\end{gather}
Now if we choose the relation,
\begin{equation}\label{dens}
\rho_{\Lambda} = - \frac{\rho_{dust}}{2} = - \rho_{SM} < 0, 
\end{equation}
the  above set of equations will be equivalent to the Eqs. \eqref{r0041}-\eqref{3ricci41}, and consequently leads to the same solution which is 
the G\"{o}del universe, given in the Cartesian coordinates as,
\begin{equation}\label{ds022}
d{s^2} = a^2 ({dt} - e^x dy)^2 - a^2 dx^2 -\frac{a^2}{2} e^{2x} dy^2 - a^2 dz^2, \\
\end{equation}
where $ a^2 = -\frac{1}{2\Lambda}$. This is the form of the metric which was originally introduced by G\"{o}del himself. 
The above form  written already in the $1+3$ form, clearly indicates a uniform gravitomagnetic 
field ${\bf B}_g =\frac {\sqrt{2}}{a^3} {\hat z}$. In terms of the cosmological constant it could be written as follows
\begin{equation}\label{ds0222}
d{s^2} = ({dT} - e^{\sqrt{2 |\Lambda|}X}dY)^2 -  dX^2 -\frac{1}{2} e^{2 \sqrt{2 |\Lambda|}X} dY^2 - dZ^2,
\end{equation}
showing clearly the flat space limit $|\Lambda| \rightarrow 0$. The metric \eqref{ds022} could also be written in the cylindrical coordinates of the form \eqref{ds02}, as follows,
\begin{equation}\label{ds033}
d{s^2} = [{dt} - 2\sqrt{2}a \sinh^2 (\frac{r}{2a}) d\phi]^2 - d r^2 - dz^2 - a^2 \sinh^2(\frac{r}{a}) d\phi^2.
\end{equation}
This form clearly indicates the regular flat space behavior near the axis ($ r \rightarrow 0$). Obviously in the second version of the spacetime source, 
we have a two-component fluid source including a dust 
component and a {\it negative density} dark fluid component ({\it negative} cosmological constant),
where the corresponding densities satisfy the first equation in \eqref{dens}, or equivalently $\Lambda = - 4\pi \rho_{dust}$. In other words in this second 
choice for the source of the spacetime, 
the requirement of having a dark fluid component, {\it automatically} results in a negative cosmological constant. It is also interesting that the relation between $\Lambda$
and $\rho_{dust}$ is
just the {\it opposite} of what we had in the case of Einstein static universe.
%%%%%%%%%%%%%%%%%%%%%%%%%%%%%%%%%%%%%%%%%%%%%%%%%%%%%%%%%%%%%%%%%%%%%%%%%%%%%%%
\subsection{Stationary spacetimes with non-vanishing $E_g$ and $B_g$: de Sitter-NUT spacetime}
Obviously keeping both fields $E_g$ and $B_g$ will leave us with more degrees of freedom, and specially one could look for {\it stationary} 
axially or cylindrically symmetric spaces. 
These are equivalent to the stationary, axially or cylindrically symmetric solutions of the Eqs. \eqref{r00}-\eqref{3ricci} with a single dark fluid 
source ($p_\Lambda = - \rho_\Lambda = constant$), which take the following forms,
\begin{gather} 
\nabla \cdot \textbf{E}_g= \frac{1}{2} h B^2_g+E^2_g + {8\pi} \rho \label{r006} \\  
\nabla \times  (\sqrt{h}\textbf{B}_g)=2 \textbf{E}_g \times (\sqrt{h}\textbf{B}_g) \label{r016} \\
{^{(3)}}P^{\mu\nu} = -{E}_g^{\mu;\nu}+\frac{1}{2}h(B_g^\mu B_g^\nu - B_g^2 \gamma^{\mu\nu})+ {E}_g^\mu E_g^\nu+ 
{8\pi} \rho \gamma^{\mu\nu} \label{3ricci6}.
\end{gather}
The axisymmetric solutions of the above equations for both positive and negative densities (cosmological constant) have already been discussed extensively 
in the literature \cite{Pleb, Podol}. Cylindrically symmetric cases are studied in \cite {Krasinski, Santos, Mac}. 
As expected, stationary exact solutions of the above equations 
contain a large family, so here, as in the previous sections, we only consider one specific solution as the family's representative. To  have the simplest solution in terms 
of the number of parameters, we look for a 2-parameter axially symmetric solution of the above equations and that is the de Sitter-(pure)NUT solution, which in a 
Schwarzschild-type coordinate system is given by,
\begin{gather}
ds^2 = \frac{F(r)}{r^2 + l^2} (dt - 2l \cos \theta d\phi)^2 - \frac{r^2 + l^2}{F(r)} dr^2 -(r^2 +l^2) \left( d\theta^2 + \sin^2 \theta d\phi^2\right) \label{NUT} \\
F(r) = r^2 - l^2 + \Lambda (l^4 - 2l^2r^2 - \frac{r^4}{3})
\end{gather} 
where for $l=0$ it reduces to (anti-)de Sitter spacetime and for $\Lambda =0 $ to the pure NUT spacetime which is the spacetime of a massless gravitomagnetic monopole \cite{MN, MN2}.
Its gravitoelectromagnetic fields are given by
\begin{gather}
E_g^r = -\frac{r(\frac{\Lambda}{3}r^4 + 3 \Lambda l^4 + \frac{2}{3} \Lambda r^2 l^2 -2l^2)}{(r^2+l^2)^2}\\
B_g^r = -2l \frac{F(r)^{1/2}}{(r^2+l^2)^{3/2}}.
\end{gather} 
It is noted that despite the apparent axial symmetry of the spacetime metric, its gravitoelectromagnetic fields are spherically symmetric \cite{MN}. 
This interesting feature is also demonstrated in the scalar invariants of the space, for example in its Kreschtmann invariant which is given by
\begin{equation}
\begin{array}{l}
K = \frac{8}{3\left(l^{2}+r^{2}\right)^{6}} \left(33 l^{12} \mathrm{\Lambda}^{2}+\mathrm{\Lambda}^{2} 
\mathrm{r}^{12}-6 l^{10} \mathrm{\Lambda}\left(8+79 \mathrm{\Lambda r}^{2}\right)
+6 l^{2} \mathrm{r}^{6}\left(-3+\mathrm{\Lambda}^{2} \mathrm{r}^{4}\right)+\right. \\
\left.9 l^{8}\left(2+80 \mathrm{\Lambda r}^{2}+55 \mathrm{\Lambda}^{2} \mathrm{r}^{4}\right)-6 l^{6} 
\mathrm{r}^{2}\left(45+2 \mathrm{\Lambda r}^{2}\left(60+\mathrm{\Lambda r}^{2}\right)\right)+3 l^{4} 
\mathrm{r}^{4}\left(90+\mathrm{\Lambda r}^{2}\left(16+5 \mathrm{\Lambda r}^{2}\right)\right)\right).
\end{array}
\end{equation}
Obviously it reduces to the Kreschtmann invariants for de Sitter ($l=0$) and pure NUT ($\Lambda=0$) spacetimes.

%%%%%%%%%%%%%%%%%%%%%%%%%%%%%%%%%%%%%%%%%%%%%%%%%%%%%%%%%%%%%%%%%%%%%%%%%%%%%%%
%%%%%%%%%%%%%%%%%%%%%%%%%%%%%%%%%%%%%%%%%%%%%%%%%%%%%%% massless
\section{Summary and discussion}
We have shown that the simplest static and stationary single and two component perfect fluid solutions of Einstein field equations, which all include 
a dark component  with EOS $p= - \rho$ ( acting as a cosmological constant), could be categorized in terms of their gravitoelectromagnetic fields. 
Apart from the stationary de Sitter-NUT solution, all the other  solutions share the same flat space limit as $|\Lambda| \rightarrow 0$. While the solutions with a double 
fluid source are given in the 
coordinate system comoving with the non-dark component, those with the single-component dark fluid are given in the  non-comoving frames. 
We treated the cosmological term as a perfect (dark) fluid with EOS $p=-\rho$, because it is only in this way that one could justify and interpret
the anisotropic feature of de Sitter-type solutions \eqref{desit3-2} and \eqref{cyldes}, in which $\Lambda$ is the only parameter. When we transform to the 
comoving frame, the anisotropic expansion in the dynamical form of
these spacetimes is traced back to the dark fluid's 3-velocity, through which
a preferred direction is inferred \cite{NKR}. In other words one could not simply identify
the geometric (cosmological constant) term $\Lambda g_{ij}$  with a
perfect fluid with the EOS $p=-\rho$, on the basis that their contribution to the energy-momentum tensor in EFEs is equivalent \cite{Schmidt}. In this way we are
ignoring the vital role of the fluid's velocity in dictating anisotropic expansion in the corresponding de Sitter-type spacetimes \cite{NKR}. 
Finally the above results could be  summarized in the following table.
\begin{table}[h]                           
\label{tab:msg1}\centering
\par
%\begin{adjustbox}{}
    \begin{tabular}{|c|c|c|}
    \hline
                 &    $E_g$ = 0   &     $E_g$ $\neq$ 0         \\ 
\hline
     $B_g$ = 0   & \makecell {Einstein Static universe \\  $\Lambda =  4\pi \rho_{dust}  > 0$ \\ dust comoving frame } &  
     \makecell {(anti-)de Sitter(-type) spacetimes\\  $\Lambda < 0 \; {\rm or} \; \Lambda > 0$  \\ non-comoving frame }    \\ \hline          
   $B_g$ $\neq$ 0& \makecell {G\"{o}del universe \\  $\Lambda = - 4\pi \rho_{dust} < 0$ \\ dust comoving frame } & 
   \makecell {de Sitter-(pure) NUT space \\ $\Lambda < 0 \;{\rm or}\; \Lambda > 0$ \\ non-comoving frame } \\ 
   \hline

    \end{tabular}
    
  %  \end{adjustbox}
 %\caption{}                               
\end{table}

\pagebreak
%%%%%%%%%%%%%%%%%%%%%%%%%%%%%%%%%%%%%%%%%%%%%%%%%%%%
\section *{Acknowledgments}
M. N-Z thanks University of Tehran for supporting this project under the grants provided by the research council. 
%%%%%%%%%%%%%%%%%%%%%%%%%%%%%%%%%%%%%%%%%%%%%%%%%%%%%%

\end{document}